\documentclass[aps,prb,amsmath,amssymb,twocolumn]{revtex4-2}
\usepackage[T1]{fontenc}
\usepackage[utf8]{inputenc}
\usepackage{graphicx}
\usepackage{bm}
\usepackage{mathptmx}
\usepackage{epstopdf}
\usepackage{booktabs}
\usepackage{multirow}
\usepackage{hhline}
\usepackage{subfigure}
\usepackage{tabularx}

\usepackage{commath}
\usepackage{bm}
\usepackage{siunitx}
\begin{document}
\title{Origin of unexpected weak Gilbert damping in the LSMO/Pt bilayer system}
\author{Pritam Das$^1$, Pushpendra Gupta$^2$, Seung-Cheol Lee$^3$}
\author{Subhankar Bedanta$^2$}
\author{Satadeep Bhattacharjee$^1$}
\email{s.bhattacharjee@ikst.res.in}
\affiliation{$^1$ Indo-Korea Science and Technology Center (IKST), Bangalore, India\\
$^2$ Laboratory for Nanomagnetism and Magnetic Materials, School of Physical Sciences, National Institute of Science Education and Research (NISER),
An OCC of Homi Bhabha National Institute (HBNI), Jatni 752050, India\\
$^3$ Electronic Materials Research Center, Korea Institute of Science $\&$ Technology, Korea}

\begin{abstract}
This study presents a first-principles and semiclassical analysis of the puzzling observation that a La$_{0.7}$Sr$_{0.3}$MnO$_3$ (LSMO) thin film exhibits \emph{larger} Gilbert damping than an LSMO/Pt bilayer, contrary to conventional spin-pumping expectations. Density‐functional theory with Wannier interpolation yields an intrinsic damping of $\alpha_{\mathrm{int}}^{\mathrm{LSMO}}\!\approx\!1.4\times10^{-3}$, supporting an extrinsic origin of the high experimental value. Guided by the self-induced inverse spin Hall effect (ISHE) demonstrated in LSMO [Gupta \textit{et al.}, Phys.\ Rev.\ B \textbf{109}, 014437 (2024)], we argue that the large spin Hall angle $|\theta_{\mathrm{SH}}|\simeq 0.093$ and low longitudinal conductivity of LSMO enable an efficient conversion of spin current to charge current boosting the effective damping. In the LSMO/Pt heterostructure the Pt cap shunts the charge current, raising $\sigma_{xx}$ and reducing the interfacial $|\theta_{\mathrm{SH}}|$ to~0.007. A Valet–Fert analysis for layer-resolved \textit{ab-initio} spin accumulation gives the Pt spin-diffusion length and a non-negligible antidamping SOT coefficient, qualitatively accounting for the observed damping reduction under current bias. The seemingly anomalous damping hierarchy is thus reconciled without invoking additional interfacial mechanisms. The distinct length scales governing spin‐pumping normalization, namely, the short absorption depth relevant to self-pumping in a single LSMO film versus the full magnetic thickness applicable to an LSMO/Pt bilayer, are crucial in this context. This observation suggests a practical design strategy: by simultaneously tuning the spin Hall-to-longitudinal conductivity ratio and the spin‐diffusion length, one can engineer heterostructures with minimized magnetic losses for spin‐orbitronics applications.
\end{abstract}
\maketitle
\section{Introduction}
In recent years, there has been a lot of interest in spintronic materials because of their potential applications in next-generation information storage and processing technology \cite{Bader2010, Zutic2004, Wolf2001}. One of the most fascinating phenomena in this subject is the interplay of ferromagnetic and non-magnetic materials, which results in emergent effects including spin pumping and the inverse spin Hall effect (ISHE) \cite{Saitoh2006, Kajiwara2010,ando2011inverse}. The recent study by Gupta et al. observed the rare but significant coexistence of anti-damping and spin pumping in manganite and heavy metal heterostructures \cite{Gupta2021}. This experimental evidence of anti-damping and ISHE in the La$_{0.67}$Sr$_{0.33}$MnO$_3$ (LSMO)/Pt bilayer system advances our understanding of the intricate interactions between ferromagnets and heavy metals and sets the stage for theoretical investigations into the underlying mechanisms.

Various works explore the relationship between high damping in materials, faster magnetization switching processes, and the requirement for higher spin currents: For example, Zhu \textit{et al.}~\cite{Zhu2020Threshold} reported that higher damping materials require greater current densities for magnetization switching, while the work of Mitani \textit{et al.}~\cite{Mitani2011Spin-transfer} highlights that while high damping improves thermal stability, it demands higher spin currents, particularly in nanopillar devices. Liu \textit{et al.}~\cite{Liu2023Dependence} showed that increasing damping in Gd-based alloys broadens the pulse fluence range for thermally induced switching and lowers the required Gd concentration. Park \textit{et al.}~\cite{Park2014Macrospin} demonstrated via macrospin modeling that high damping improves sub-nanosecond switching reliability but requires careful spin-orbit torque optimization. 

The LSMO/Pt bilayer system possesses unique properties due to its components. LSMO is a well-known colossal magnetoresistance material that exhibits ferromagnetic order with a high Curie temperature, making it an attractive choice for spin injection \cite{Andres2003, Urushibara1995, Park1998}. Platinum, on the other hand, is a heavy metal with strong spin-orbit coupling, which facilitates efficient spin-to-charge current conversion via the ISHE \cite{Sagasta2016, Hoffmann2013}. The interaction of these materials in heterostructures provides an ideal environment for studying spin dynamics and spin-charge conversion. The simultaneous findings of anti-damping and the ISHE in this system indicate a complex interaction between spin pumping from the ferromagnetic LSMO layer and spin current absorption and conversion in the Pt layer \cite{Huang2012}.

Previous studies have shown that when a heavy metal (HM) is deposited adjacent to a ferromagnet (FM), the FM generates a spin current during spin pumping, which is dissipated in the HM. This spin current absorption in the HM leads to the relaxation of spins in the FM, thereby enhancing the effective Gilbert damping. Factors such as proximity-induced magnetization in the HM, interface roughness, etc., can also enhance the damping in the system near the HM \cite{Swindells2022, Swindells2021}. This increased damping can limit the application of such devices in high-speed dynamics. However, there are instances where, instead of an enhancement, a reduction in effective damping, known as anti-damping, has been observed \cite{Gupta2021, Behera2016}. Anti-damping, defined as the reduction in effective Gilbert damping in FM/HM bilayers compared to single FM films, suggests efficient spin injection and absorption at the interface, enabling the utilization of such devices without compromising their high-speed dynamical properties \cite{Lattery2018}.

A comprehensive theoretical framework is required to clarify the microscopic mechanisms behind anti-damping events. Density Functional Theory (DFT) simulations are widely used for studying the electronic structure and spin-dependent interactions at the atomic level \cite{Tserkovnyak2002, Tserkovnyak2005}. DFT can help us understand the electrical conductivity and spin Hall conductivity of LSMO and Pt, as well as their interface. The efficiency of spin injection and subsequent spin-to-charge conversion processes are closely related to the electrical and spin Hall conductivities of the thin films. These computations can also provide insights into the role of interfacial states and the impact of structural and electronic features on observed spintronic phenomena.

In this work, we aim to bridge the gap between experimental observations and theoretical understanding of the LSMO/Pt bilayer system. By employing DFT calculations for LSMO, Pt, and LSMO/Pt heterostructures, we investigate their electrical and spin Hall conductivities. Additionally, to place our results in a broader context, we conduct a comparative study on the Co/Pt system, a benchmark for spin-pumping studies.

Our study proposes the experimental observation of higher damping pure LSMO can be related to the higher $\theta_{SH}$ in the LSMO films compared to the Pt/LSMO system where the presence of the metallic Pt layer increases the longitudinal conductivity (\(\sigma_{xx}\)), resulting in a much lower \(\theta_{SH}\).
 
In contrast, the Co/Pt system exhibits a different behavior. The longitudinal conductivity and spin Hall conductivity in Co and Co/Pt do not show significant changes due to the addition of Pt. The spin Hall angle in Co/Pt is observed to be larger than in single Co films, indicating that the presence of Pt enhances the spin pumping effect and increases the Gilbert damping. This behavior is consistent with the well-known mechanisms in Co/Pt systems, where the enhanced spin current absorption in Pt leads to increased damping in the ferromagnetic layer.

Our findings indicate that the mechanisms responsible for the observed damping behavior in LSMO and LSMO/Pt do not appear in the Co and Co/Pt systems. In LSMO, the self-induced inverse spin Hall effect (ISHE) plays a crucial role, leading to higher damping due to efficient spin-to-charge conversion within the material. In LSMO/Pt, the interplay between the reduced \(\theta_{SH}\) and the larger spin diffusion length results in a notable anti-damping contribution, balancing the overall damping. This intricate interplay of spin dynamics is not observed in Co/Pt, where the damping behavior is primarily governed by spin current absorption in the Pt layer. 
\begin{figure}
\includegraphics[width=1.0\columnwidth]{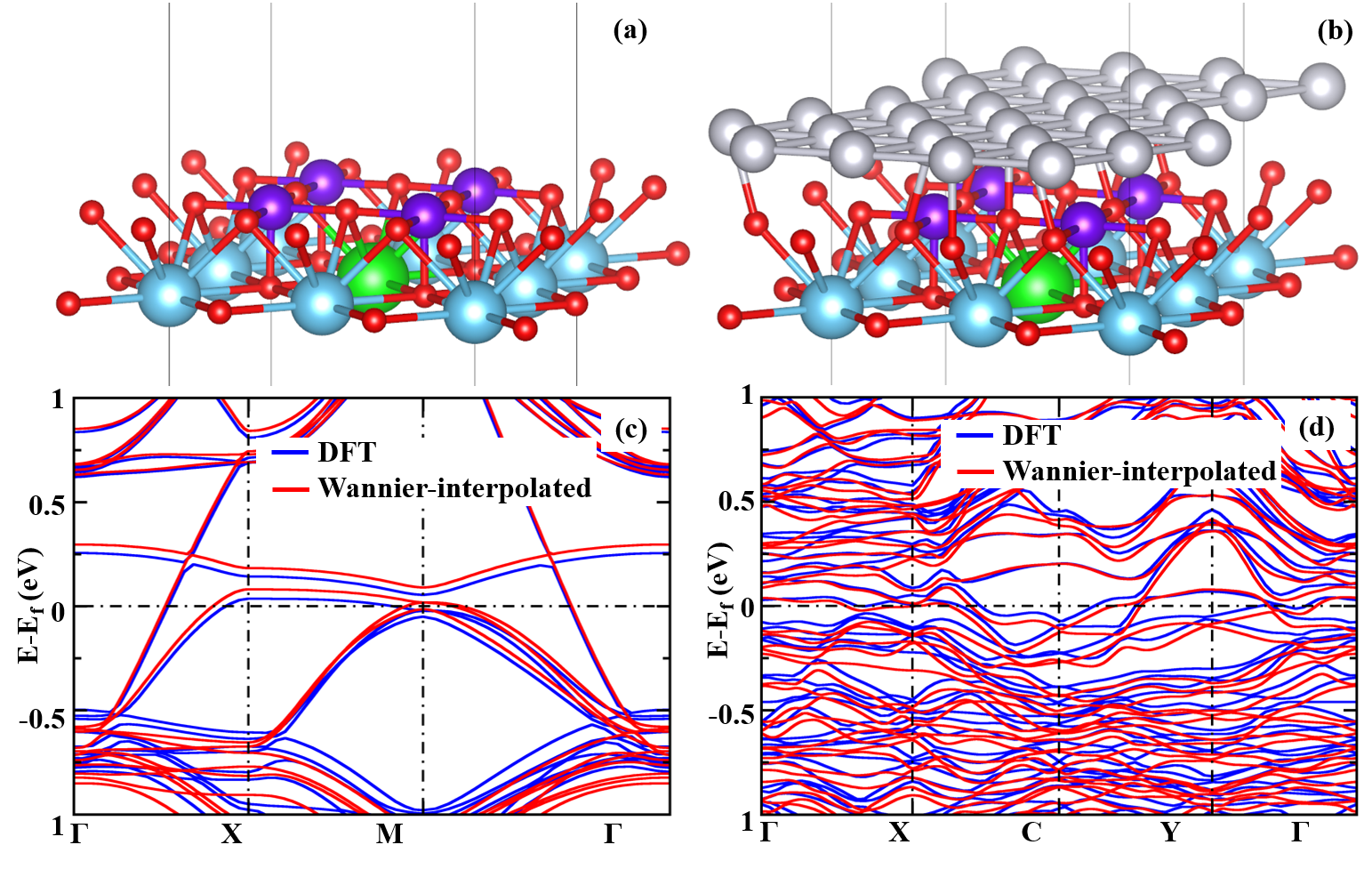}
\caption{Structural models and electronic band structures of LSMO and LSMO/Pt films. The top panel (a\&b) corresponds to the structural models while bottom panels displays DFT band structure fitted with Wannier90 interpolated band structure for both LSMO (c) and LSMO/Pt (d).}
\label{Fig1}
\end{figure}
\section{Computational methodology}
\subsection{First-Principles DFT Calculations}
The ground state \textit{ab initio} calculations were carried out in the framework of DFT implemented in the Vienna ab initio Simulation Package (VASP)~\cite{Kresse1996}. The electronic band structure was calculated within the generalized gradient approximation (GGA) exchange-correlation functional of Perdew-Burke Ernzerhof (PBE)~\cite{Perdew1996} with spin-orbit coupling (SOC) terms. All energies converged within a cutoff of 450 eV. The conjugate gradient algorithm was used for structural optimization. The convergence criteria for energy and force were $10^{-6}$ eV and $0.05$ eV/\AA, respectively. Once the ground state is obtained, the transport properties were calculated using the WANNIER90 package~\cite{pizzi2020wannier90}. The longitudinal charge conductivity was calculated using the BOLTZWANN module~\cite{pizzi2014boltzwann}. To simulate the LSMO and LSMO/Pt films we considered the slab model as shown in Fig.\ref{Fig1}. The LSMO film was constructed with two monolayers of bulk LSMO (001) surface while LSMO/Pt film was constructed using the heterostructure of one monolayer of Pt (111) surfaces stacked on two monolayers of LSMO (001). A vacuum of 10\AA~ was used. Dipole corrections were performed along the z-direction to eliminate the unwanted electric field effects.
\subsection{Wannier Interpolation and Transport Properties}
To understand the relative magnitude of damping and anti-damping torques in LSMO and LSMO/Pt systems, we compared the spin Hall angles (\(\theta_{SH}\)) in these two systems. The spin Hall angle is a crucial parameter that indicates the efficiency of the conversion from charge current to spin current, thereby influencing the strength of the spin pumping as well as spin-orbit torques~\cite{ketkar2023enhanced,gambardella2011current,manchon2019current,guimaraes2020spin}.  This finding is significant because it demonstrates a unique mechanism for spin current generation and spin-to-charge conversion that does not rely on spin currents generated in adjacent layers, as is typically seen in traditional measurement methods for the spin Hall effect.

The spin Hall angle is given by the following equation~\cite{Qu2014Self-consistent},
\begin{equation}
\theta_{SH} = \frac{2e}{\hbar} \frac{\sigma_{xy}^{SHC}}{\sigma_{xx}}
\label{SH}
\end{equation}
Where \(\sigma_{xy}^{SHC}\) is the spin Hall conductivity (SHC) and \(\sigma_{xx}\) is the longitudinal charge conductivity.

The intrinsic SHC was calculated by fitting DFT Hamiltonian to an effective Wannier Hamiltonian using Wannier90 approach and by using Kubo's formula~\cite{qiao2018calculation},
\begin{equation}
\sigma_{xy}^{{SHC}} (E) = -\frac{e^2}{\hbar} \frac{1}{V N} \sum_{n k} \Omega_{n \mathbf{k}, x y}^{\text {spin,z}}(E)f_{nk}
\end{equation}

The spin Berry curvature \(\Omega_{n,xy}(\mathbf{k})\) can be obtained using the following equation~\cite{qiao2018calculation},

\begin{equation}
\Omega_{n \mathbf{k}, x y}^{\text {spin,z}}(E)=\hbar^2 \sum_{m \neq n} \frac{-2 \operatorname{Im}\left[\left\langle\psi_{n \mathbf{k}}\right| \frac{2}{\hbar} \hat{j}_x\left|\psi_{m \mathbf{k}}\right\rangle\left\langle\psi_{m \mathbf{k}}\right| \hat{v}_y\left|\psi_{n \mathbf{k}}\right\rangle\right]}{\left(\epsilon_{n \mathbf{k}}-\epsilon_{m \mathbf{k}}\right)^2-(E+i \eta)^2}
\end{equation}
Here $\hat{j}_x^s$ is the spin current operator and is given by $\hat{j}_x^s=\frac{\hbar}{2}\left\{\sigma_z, v_x\right\}$ which represents the spin current flowing in the x-direction with spin polarization along the z-axis. \(\epsilon_{nk}\) and \(\epsilon_{mk}\) are the eignevalues of the Bloch states \(| nk \rangle\) and \(| mk \rangle\), respectively. The above equation for  \(\Omega_{n \mathbf{k}, x y}^{\text {spin,z}}(E)\) calculates the spin Berry curvature by summing over all possible virtual transitions between an occupied band \(n\) and other bands \(m\) (excluding \(n\) itself). It is therefore a measure of  spin and charge currents mediate these transitions, weighted by the inverse square of the energy differences. The spin current operator \(\hat{j}_x^s\) and the velocity operator \(\hat{v}_y\) capture the essential physics of spin and charge transport in the material, respectively.
\subsection{Gilbert Damping via Torque‐Correlation}
The intrinsic Gilbert damping was calculated within Kambersky's torque-correlation model which is given by~\cite{guimaraes2019comparative},
\begin{equation}
\alpha_{int}=\frac{2\gamma}{M_s \pi N} \sum_k\operatorname{Tr}\left[\operatorname{Im} G\left(k,E_{\mathrm{F}}\right) \hat{T}^{-} \operatorname{Im} G\left(k,E_{\mathrm{F}}\right) \hat{T}^{+}\right]
\end{equation}
Where, $\hat{T}^{ \pm}=\left[\hat{S}^{ \pm}, \hat{\mathrm{H}}_{\mathrm{SO}}\right]$ are the torque-operators. $G\left(k,E_{\mathrm{F}}\right)=[E+i\eta-H(k)]^-1$ is the Green's function at Fermi energy $E_F$. The Hamiltonian ${H} = {H}_0 + {H}_{xc} + {H}_{SOI}$ describes the electronic structure of the system. The paramagnetic band structure is described by $H_0$, and ${H}_{\text {exc }}$ describes the effective local electron-electron interaction, which is treated within a spin-polarized semi-local Kohn-Sham exchange-correlation functional approach. ${H}_{\mathrm{SO}}$ is the spin-orbit Hamiltonian and $\hat{S}^{ \pm}$ are the spin operators. $N$ is the number of k-points, while $M_s$ is the saturation moment and $\gamma$ is the gyromagnetic ratio. $\eta$ is the broadening, which was set to be 10 meV in our calculation.
\section{Results and discussion}
\subsection{Electronic Structure and Wannier Validation}
Figures 1(b) and 1(d) present the electronic band structures of LSMO and LSMO/Pt, respectively. The band structures reveal that LSMO retains its metallic nature with bands crossing the Fermi level, while LSMO/Pt shows more complex band hybridization due to the interaction with Pt layers. This hybridization can influence the spin-orbit coupling and subsequently the spin-Hall effect~\cite{Huang2012Transport,Zhang2015Reduced,Alotibi2021Enhanced,Lyalin2021Spin-Orbit}. The figure also demonstrates a successful Wannier interpolation in both cases.

\subsection{Intrinsic damping }
It should be noted here that LSMO being a half-metal is expected to have smaller intrinsic Gilbert damping as can be understood either from Kambersky's torque-torque model~\cite{kambersky2007spin,liu2009origin} or from the work of Gilmore \text{et al.}~\cite{gilmore2007identification}. 
 We obtain the intrinsic damping to be 0.0014 for the LSMO, while it is 0.012 for Pt/LSMO. As LSMO is half metal, there is no spin-flip scattering for the minority spin electrons due to the reduction in the available states, which leads to fewer scattering events that contribute to energy dissipation, thereby reducing the intrinsic damping.
Therefore, the appearance of a relatively larger damping in LSMO films can only be rationalized by an extrinsic source such as self-induced ISHE.

It is important to acknowledge that while LSMO demonstrates half-metallicity as temperature approaches zero, its spin polarization is observed to decrease due to factors such as structural imperfections, non-zero temperatures, and defects introduced during growth. Point-contact Andreev reflection and spin-resolved photoemission studies generally report spin polarization values of approximately $P\simeq 0.75{-}0.85$ at 300 K for epitaxial films on $\mathrm{SrTiO}_3$. In contrast, polycrystalline samples may exhibit values below 0.6.~\cite{bowen2003nearly,bibes2007oxide}

\subsection{Longitudinal charge ($\sigma_{xx}$) and spin Hall ($\sigma_{SHC}$) conductivities}
The longitudinal conductivity (\(\sigma_{xx}\)) and spin Hall conductivity (\(\sigma_{xy}\)) for both systems are depicted in Figure 2. The \(\sigma_{xx}\) plots (Figures 2a and 2c) highlight a stark contrast between the two systems. For LSMO, the small longitudinal conductivity is observed at zero chemical potential ($\mu=0$), indicating limited conducting channels. In contrast, the presence of Pt in LSMO/Pt significantly enhances \(\sigma_{xx}\) at zero chemical potential due to the additional pathways for charge transport which can be further understood from the Fig.\ref{Fig1} where it is clear that Pt introduces additional states near the Fermi energy facilitating better charge transport. As can be seen from the figure, the longitudinal conductivity (\(\sigma_{xx}\)) for LSMO is very small at \(\mu = 0\), indicating poor charge transport properties near the Fermi level (\(E_F\)). This suggests a low density of states (DOS) at the Fermi level which is consistent with the electronic band structure shown in Fig.\ref{Fig1} which depicts a single band crossing the Fermi level. However, it is important to note that at higher chemical potentials, LSMO has larger $\sigma_{xx}$. 

It may be noted here that, although the interface between LSMO and Pt may exhibit finite roughness and a 1.3\% lattice mismatch, these effects are not expected to overshadow the pronounced disparity in longitudinal conductivities. The additional scattering arising from interface imperfections could modify the transport behavior, but the substantial conductivity difference of over an order of magnitude suggests that our main conclusions remain valid.

In order to calculate the longitudinal charge conductivity, we used a relaxation time, $\tau=10 \text{fs}$. We also further investigate the significance of the \(\tau\) in determining the absolute values of conductivity. Assuming that the dominant scattering mechanism is acoustic phonon scattering, the scattering rate due to the acoustic deformation potential (ADP) is expressed as~\cite{lundstrom2002fundamentals,mandia2021ammcr},
\begin{equation}
\frac{1}{\tau_{\text{ADP}}(E)} = \frac{\pi D_A^2 k_B T}{\hbar \rho v_s^2} D(E)
\end{equation}

where \(D_A\) is the acoustic deformation potential, \(D(E)\) represents the density of states (DOS) at energy \(E\), \(k_B\) is the Boltzmann constant, \(T\) is the temperature, \(\hbar\) is the reduced Planck constant, \(\rho\) is the mass density, and \(v_s\) is the sound velocity. 

Assuming that \(\rho\), \(v_s\), and \(T\) are constant for both LSMO and LSMO/Pt, we derived the ratio of their relaxation times as,
\[
\frac{\tau_{\text{LSMO}}}{\tau_{\text{LSMO/Pt}}} = \frac{[D_A(\text{LSMO})]^2 D_{\text{LSMO}}(E)}{[D_A(\text{LSMO/Pt})]^2 D_{\text{LSMO/Pt}}(E)}.
\]

Using the computed density of states (DOS) at the Fermi level and the acoustic deformation potentials derived from DFT calculations, we find:
\[
\frac{\tau_{\text{LSMO}}}{\tau_{\text{LSMO/Pt}}} = \frac{26.0162}{12.4112} \approx 2.095.
\]

This indicates that the relaxation time in LSMO is approximately 2.1 times longer than in LSMO/Pt. Furthermore, we recalculated the longitudinal charge conductivity for LSMO using a relaxation time of \(\tau = 21\) fs, which resulted in a conductivity of \(27738.05 \, \text{S/m}\), compared to \(13869.02 \, \text{S/m}\) calculated earlier with \(\tau = 10\) fs. Despite this increase, LSMO's conductivity remains significantly lower than that of LSMO/Pt, which is \(295584.22 \, \text{S/m}\)—approximately 11 times larger.

\begin{figure}
\includegraphics[width=1.0\columnwidth]{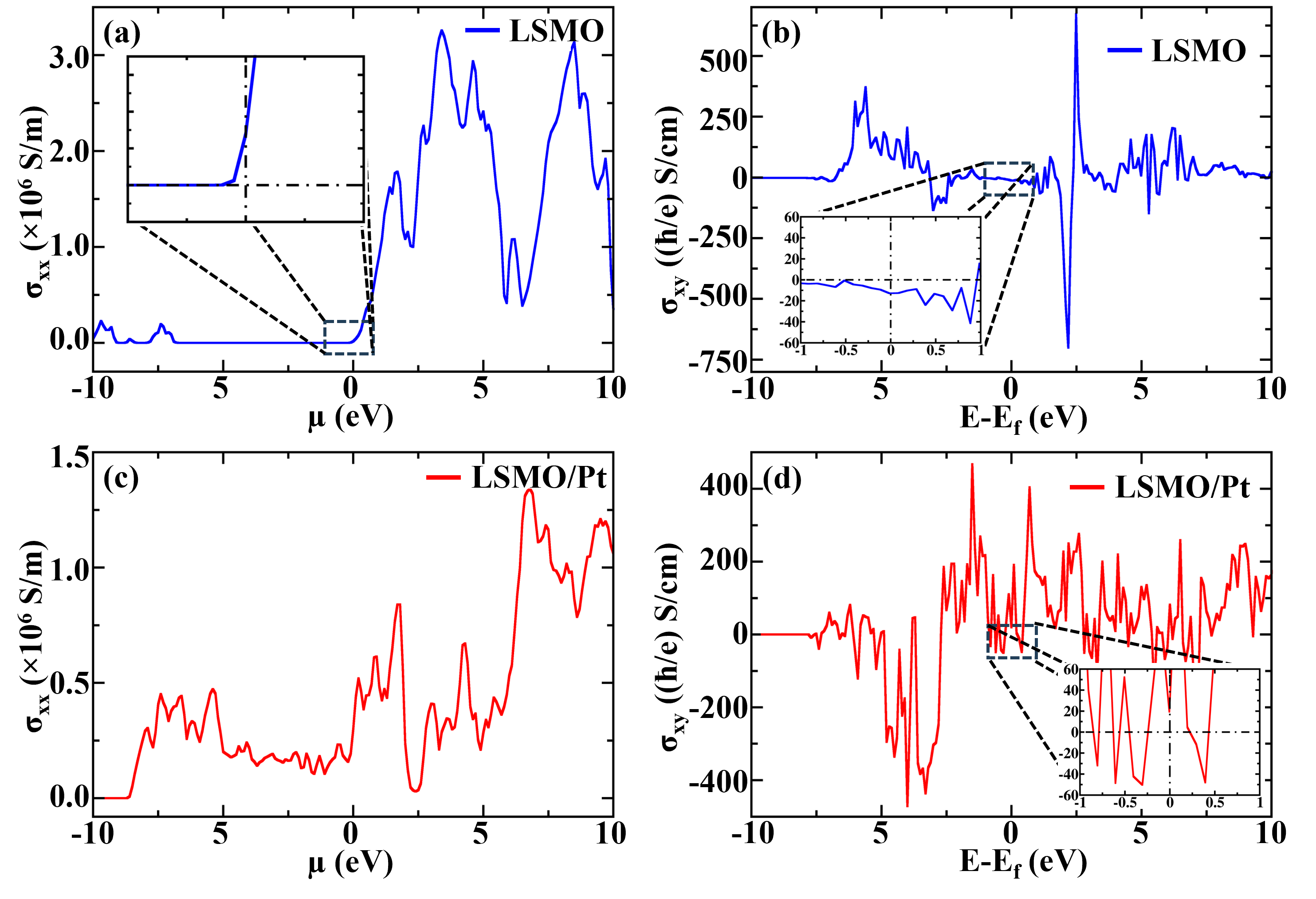}
\caption{The longitudinal charge conductivity and spin Hall conductivities. Upper panels (a\&b):LSMO, lower panels (c\&d): LSMO/Pt }
\label{Fig2}
\end{figure}

In principle, by doping to adjust the chemical potential, one can effectively tune the $\theta_{\text{SH}}$ in both LSMO and LSMO/Pt systems. It can be easily understood from the 
Fig.\ref{Fig2} that n-type doping can reduce the $\theta_{\text{SH}}$ in LSMO further. It should be noted that doping plays a crucial role in modulating the spin Hall conductivity (SHC) and spin Hall angle by altering the electronic structure and scattering mechanisms in materials~\cite{Garlid2010Electrical,Qu2018Large,Zhou2018Intrinsic,Matsuzaka2009Scanning}.

\subsection{Self-induced ISHE in LSMO and its impact on effective damping}
The spin Hall conductivities (\(\sigma_{xy}\)) for LSMO and LSMO/Pt (Figures 2b and 2d) further elucidate the spintronic properties.  For LSMO, \(|\theta_{\text{SH}}|\) is found to be 0.093  whereas for LSMO/Pt, it reduces to 0.007. This reduction is primarily attributed to the increased longitudinal conductivity (which is in the denominator of Eq.\ref{SH}) in LSMO/Pt due to Pt bands crossing the Fermi energy, which dilutes the effective spin Hall efficiency.
The observed Gilbert damping in LSMO is larger than in LSMO/Pt, which can be rationalized by considering the details of the spin-orbit torque mechanisms. It is important to note that recently Gupta \textit{et al.} reported observation of a self-induced inverse spin Hall effect (ISHE) and spin pumping~\cite{gupta2024self} within LSMO. It can be noted that similar self-induced ISHE was observed by Tsukahara
\textit{et al.}~\cite{tsukahara2014self} in permalloy at room temperature. This self-induced ISHE leads to the generation of a significant spin-pumping voltage within the LSMO film, indicating the presence of appreciable spin-to-charge conversion within the material. In the case of LSMO the absence of an external interface means there is no anti-damping mechanism from an external source (due to the injection of spin current from another material). The self-induced spin pumping effect powered by larger values of $|\theta_{\text{SH}}|$ in LSMO, therefore, leads to significant angular momentum dissipation within the material itself, contributing to larger effective damping via conversion of spin current to charge current through $\mathbf{J}_c = \theta_{SH} \left( \frac{\hbar}{2e} \right) (\mathbf{J}_s \times \mathbf{\sigma})$. Here $\mathbf{J}_c$ and  $\mathbf{J}_s$ are charge and spin current density, respectively. $\mathbf{\sigma}$ is the direction of the spin polarization~\cite{saitoh2006conversion}.

Another important point we can gather from the sign of the spin Hall angle we have computed: the sign of \(\theta_{SH}\) determines the direction of the charge current generated by the ISHE. It can be seen that \(\theta_{SH}\) for LSMO is negative. However, the absolute value \(|\theta_{SH}|\) indicates the strength of the spin-to-charge conversion. Thus, a large \(|\theta_{SH}|\) (regardless of whether \(\theta_{SH}\) is positive or negative) can result in a significant ISHE.
Recently, Bleser \textit{et al.}~\cite{bleser2022negative} demonstrated large spin-to-charge conversion with a negative spin Hall angle in thermally evaporated chromium thin films. The effective sign of \(\theta_{SH}\)  can also be influenced by the thickness of the film. The difference in the composition between the MnO$_2$ and (La,Sr)O layers in the structural model we considered (see Fig.\ref{Fig1}) therefore induces a strong SOC at the interface. This gradient can enhance the ISHE in a manner similar to FePt films as seen in recent experiments~\cite{ampuero2024self}.

In the LSMO/Pt heterostructure, on the other hand, the presence of SHE guarantees the presence of ISHE as well. The ISHE contributes to spin pumping, where spin currents from LSMO are absorbed by Pt, leading to angular momentum dissipation as mentioned above which contributes to the overall damping in addition to the intrinsic damping. 
The SHE in Pt generates spin currents in response to an applied charge current is given by $\mathbf{J}_s = \theta_{SH} \left( \frac{2e}{\hbar} \right) (\mathbf{J}_c \times \mathbf{\sigma})$
which can flow back into LSMO and exert anti-damping torques.  The presence of both mechanisms therefore competes here and balances the overall damping in the LSMO/Pt heterostructure. A small value of \(\theta_{\text{SH}}\) implies a less efficient spin-to-charge conversion. However, in the experiment,  an antidamping torque is still seen~\cite{Gupta2021}. This can be attributed to the larger spin diffusion length present in the system. 

It is worth noting that, although self-induced ISHE is prominent in the experiment if the self-induced ISHE seen in LSMO is associated with MnO$_2$ layers as the source of spin current and (La, Sr)O layers as the source of spin-orbit interaction, it also necessitates considering the inverse effect, namely the anti-damping torque due to the intrinsic SHE in the (La, Sr)O layer and the generation of spin current. It can be argued here that in ultrathin layers, the anti-damping torque, often associated with the SHE in the heavy metal, might be suppressed due to the reduced spin diffusion length in very thin HM layers. The spin current generated by SHE may not effectively reach the FM layer to induce significant anti-damping torque, especially if the thickness of the HM layer is less than the spin diffusion length. It can be noted that while a shorter spin diffusion length supports the spin-pumping via ISHE (as demonstrated by ~\cite{fache2020determination} that CoFeB/Ir with spin diffusion length of 1.3 nm has higher spin pumping effect in comparison to CoFeB/Pt with a larger spin diffusion length of 2.6 nm) as spin currents are absorbed more effectively near the interface between the ferromagnet and the non-magnetic layer, the longer spin diffusion lengths are better suited for generating effective spin-orbit torques (such as anti-damping torque). 
Another possible contribution to the absence of such an effect in the experiment could come from the magnetic moment. The anti-damping torque due to the SHE effect scales as \(\frac{1}{M_s^2}\) \cite{sinova2015spin}, while the ISHE scales as \(\frac{1}{M_s}\) \cite{ampuero2024self}. From our DFT calculations, we find the value of \(M_s\) for LSMO and LSMO/Pt films to be \(8.3 \mu_B\) and \(7.02 \mu_B\), respectively.  Finally, the efficiency parameter~\cite{sinova2015spin} might be small for LSMO.

The generation of spin-orbit torques, such as the antidamping torque, in heavy metal (HM)/ferromagnet (FM) bilayers is a complex interplay of several material and interface parameters. In the following we analyse the interface properties.
\subsection{Spin-diffusion length in Pt and antidamping SOT in LSMO/Pt}
To understand the behavior of spin transport and determine the spin-diffusion length (\(\lambda_{\text{sf}}\)) in the La$_{0.7}$Sr$_{0.3}$MnO$_3$/Pt (LSMO/Pt) system, we performed first-principles calculations on a model heterostructure. This structure consisted of two monolayers (ML) of LSMO and four monolayers of Pt (\(t_{\text{Pt}} \approx 0.8\) nm), as depicted in Fig.~\ref{Fig4} . We obtained the layer-resolved spin polarization within the Pt layers using spin-polarized Density Functional Theory (DFT). The spin polarization \(P(z)\) at a given position \(z\) along the film orientation (normal to the interface) was calculated from the spin-dependent density of states (DOS) at the Fermi energy (\(E_F\)) using the established relation~\cite{bhattacharjee2019first}:
\begin{equation}
P(z)=\frac{D^{\uparrow}\left(E_F,z\right)-D^{\downarrow}\left(E_F,z\right)}{D^{\uparrow}\left(E_F,z\right)+D^{\downarrow}\left(E_F,z\right)},
\label{eq:PL}
\end{equation}
where \(D^{\uparrow(\downarrow)}\left(E_F,z\right)\) represents the DOS for majority (minority) spin electrons at \(E_F\) in layer \(z\).

\begin{figure}
\includegraphics[width=0.96\columnwidth]{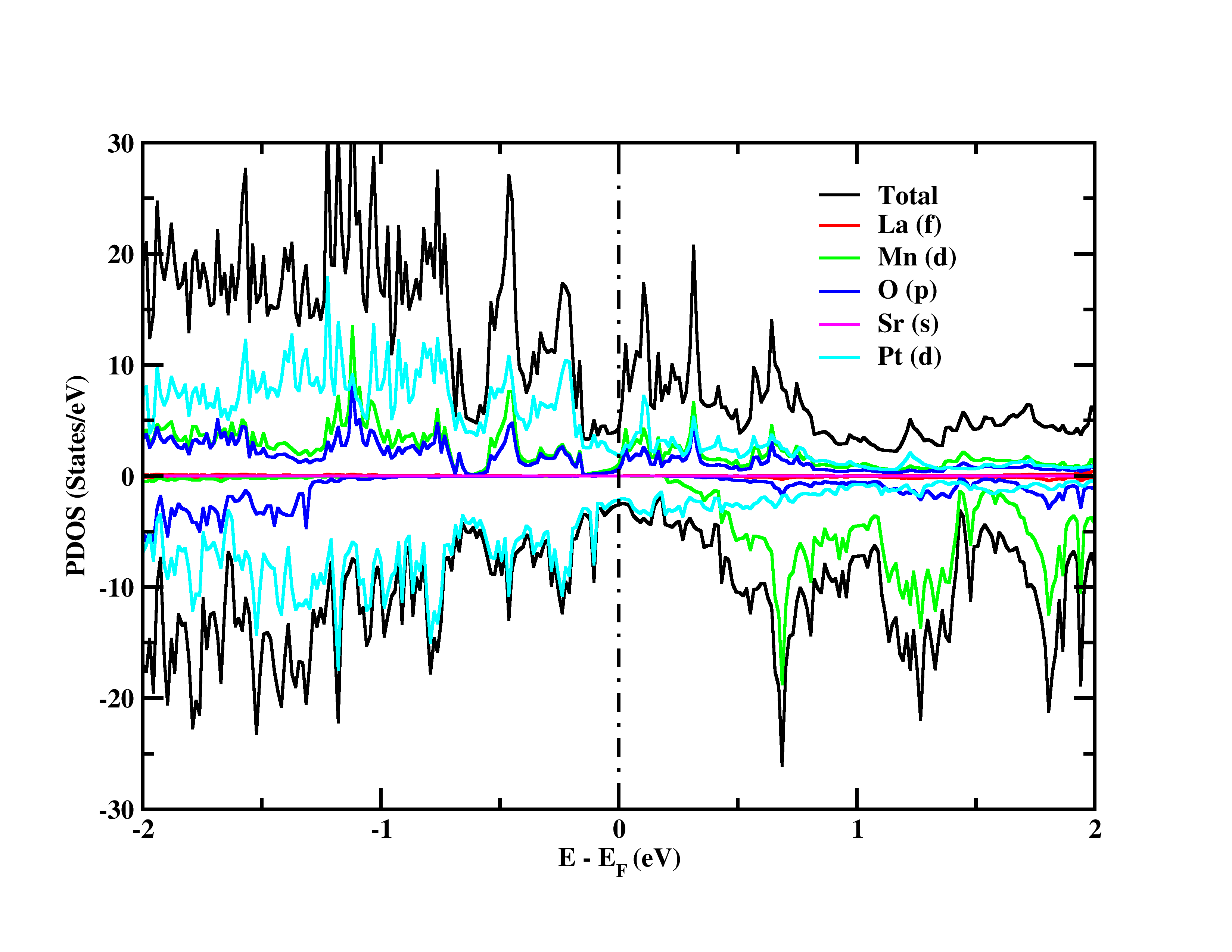}
\caption{Spin-resolved total and atom-projected partial density of states (PDOS) for the LSMO/Pt heterostructure. The Fermi level (dashed vertical line) is predominantly populated by Pt-\emph{d} states, with negligible contribution from Mn-\emph{d} orbitals. }
\label{Fig5}
\end{figure}

\begin{figure}
\includegraphics[width=0.96\columnwidth]{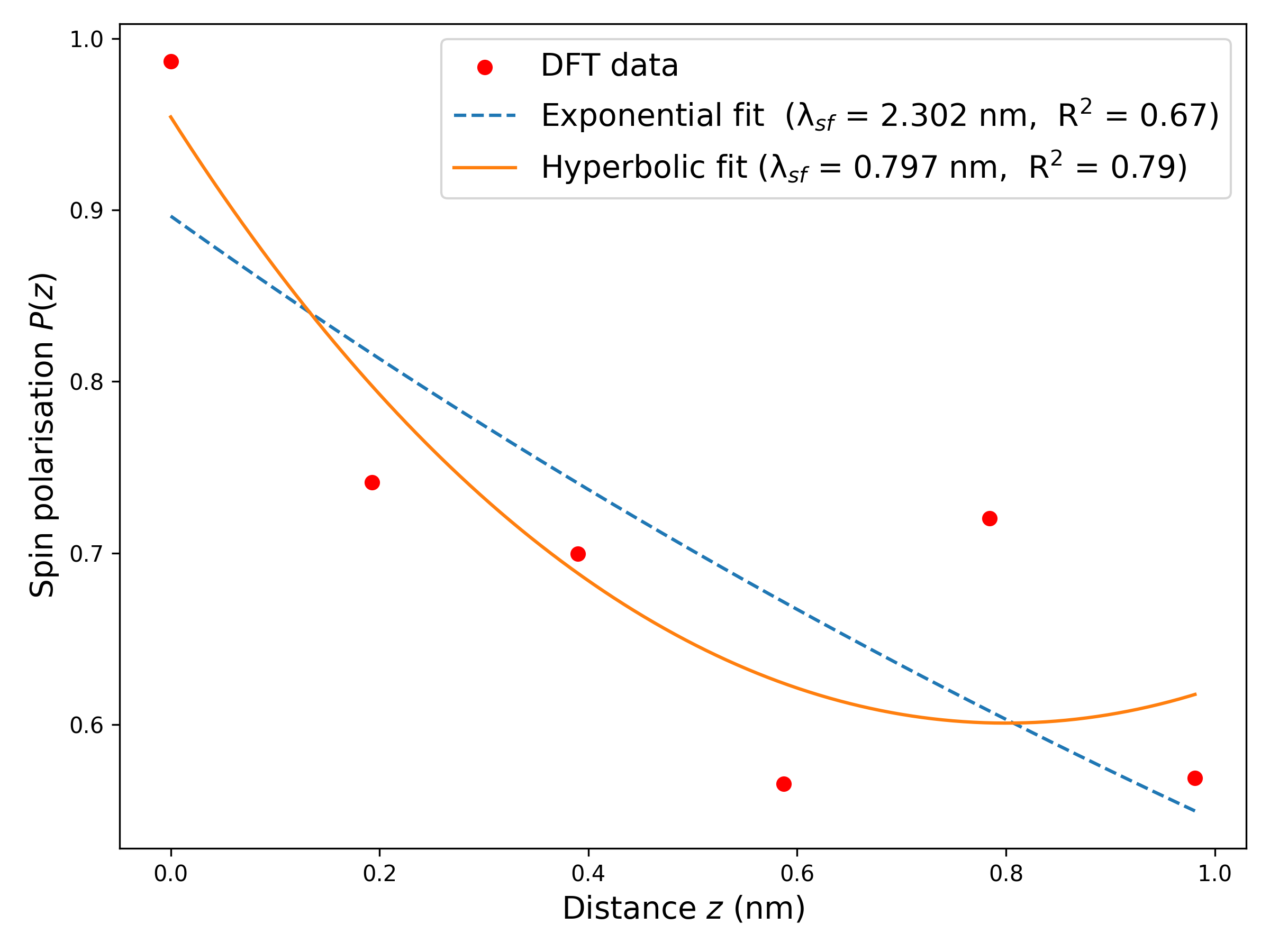}
\caption{Layer dependent spin polarization of the LSMO/Pt hetero-structure obtained by fitting the DFT-data with the Eq.\ref{eq:P_cosh} (orange line). For completeness, an exponential fit with \(P(z) \approx P_0 \exp(-z/\lambda_{\mathrm{sf}}) + C\) (blue line) is also shown.}
\label{Fig4}
\end{figure}

In Fig.\ref{Fig5}, the partial density of states (PDOS) for the LSMO/Pt system shows that near the Fermi energy, the Pt-\emph{d} bands dominate, whereas the Mn-\emph{d} states—which could otherwise introduce a heavier orbital character—are found farther from \(E_F\). This is apparent from the near-negligible Mn-\emph{d} contribution around \(E_F\) relative to the robust Pt-\emph{d} weight. According to the work of Mazin \textit{et al.},\cite{mazin1999define}, the significant discrepancy between DOS-based and transport-based spin polarizations generally occurs when both light \emph{s}-like and heavy \emph{d}-like states coexist at \(E_F\). In contrast, here only the Pt-\emph{d} orbitals are relevant near the Fermi level, suggesting that the difference between \(P_N\) (DOS-based) and \(P_{Nv^2}\) (transport-based) should be minimal. The predominance of a single orbital effectively bypasses the issues that occur with multiple orbital characters having varying Fermi velocities. This ensures the credibility of our density of states (DOS)-based qualitative analysis regarding the local spin asymmetry within the LSMO/Pt system in a reasonable manner.

Within the framework of the diffusive Valet-Fert theory~\cite{valet1993theory}, a stationary spin accumulation \(\mu_s(z)=\mu_\uparrow(z)-\mu_\downarrow(z)\) within a non-magnetic material obeys the diffusion equation:
\begin{equation}
\frac{d^{2}\mu_s(z)}{dz^{2}} =\frac{\mu_s(z)}{\lambda_{\mathrm{sf}}^{2}}.
\label{eq:valet_fert_diff}
\end{equation}
In the linear-response regime, where the induced spin accumulation is small, it is reasonable to assume that the DFT-calculated spin polarization \(P(z)\) is proportional to \(\mu_s(z)\) with slowly varying profiles along z.. The general solution to Eq.~\eqref{eq:valet_fert_diff} for a film of thickness \(t_{\text{Pt}}\), with appropriate boundary conditions (e.g., \(z=0\) at the LSMO/Pt interface and \(z=t_{\text{Pt}}\) at the Pt/vacuum interface), can be expressed in a hyperbolic cosine form~\cite{valet1993theory,nair2021spin},
\begin{equation}
P(z)=P_0 \frac{\cosh\!\bigl[(t_{\mathrm{Pt}}-z)/\lambda_{\mathrm{sf}}\bigr]}{\cosh\!\bigl(t_{\mathrm{Pt}}/\lambda_{\mathrm{sf}}\bigr)} + C,
\label{eq:P_cosh}
\end{equation}
where \(P_0\) is the spin polarization at the interface (\(z=0\)), \(\lambda_{\mathrm{sf}}\) is the spin-diffusion length, and \(C\) is an offset constant. This hyperbolic form reduces to a simpler exponential decay, \(P(z) \approx P_0 \exp(-z/\lambda_{\mathrm{sf}}) + C\), only in the thick-film limit where the film thickness \(t_{\text{Pt}}\) is much greater than the spin-diffusion length (\(t_{\text{Pt}} \gg \lambda_{\mathrm{sf}}\)).

In the experimental setup by Gupta \textit{et al.}~\cite{Gupta2021}, the Pt layer thickness was 3 nm. A direct first-principles treatment of such a thick slab would be computationally prohibitive. Therefore, our DFT calculations employed a thinner 4-monolayer Pt film (\(t_{\text{Pt}} \approx 0.8\) nm). Initially, an exponential decay model was considered for fitting the DFT-derived \(P(z)\) data, which yielded \(\lambda_{\text{sf}} \approx 2.175\) nm. However, this result (\(\lambda_{\text{sf}} \approx 2.7 \times t_{\text{Pt}}\)) indicated that the condition \(t_{\text{Pt}} \gg \lambda_{\text{sf}}\) was not met, making the exponential approximation inappropriate for our DFT slab. Therefore, we again fitted the layer-resolved DFT polarization data from the 4 ML Pt slab directly to the more general hyperbolic cosine model given by Eq.~\eqref{eq:P_cosh}. This fitting procedure yielded a spin-diffusion length of \(\lambda_{\text{sf,Pt}} \approx 0.7975\) nm (\(7.975\) Å). This value, where \(\lambda_{\text{sf,Pt}} \approx t_{\text{Pt}}\), confirms the necessity of using the hyperbolic model for our computationally constrained thin Pt slab. While this \(\lambda_{\text{sf,Pt}}\) is shorter than some bulk Pt values, it is a plausible result for an ultrathin film where interface scattering and confinement effects can significantly influence spin transport.

Using the spin-diffusion length obtained as above and employing the semiclassical model developed by Haney \textit{et al.}~\cite{haney2013current}, which provides an expression for the dimensionless damping-like torque coefficient, \(\tau_{ad}\) as,

\begin{align}
\tau_{ad} &= \theta_{\text{SH}} \frac{(1 - e^{-t_{\text{Pt}}/\lambda_{\text{sf,Pt}}})^2}{1 + e^{-2t_{\text{Pt}}/\lambda_{\text{sf,Pt}}}} \notag \\
&\quad \times \frac{|\tilde{G}^{\uparrow\downarrow}|^2 + \text{Re}[\tilde{G}^{\uparrow\downarrow}] \tanh^2(t_{\text{Pt}}/\lambda_{\text{sf,Pt}})}%
{|\tilde{G}^{\uparrow\downarrow}|^2 + 2\text{Re}[\tilde{G}^{\uparrow\downarrow}] \tanh^2(t_{\text{Pt}}/\lambda_{\text{sf,Pt}}) + \tanh^4(t_{\text{Pt}}/\lambda_{\text{sf,Pt}})}
\end{align}
Where the dimensionless scaled spin-mixing conductance, \(\tilde{G}^{\uparrow\downarrow}\), is defined through $\tilde{G}^{\uparrow\downarrow} = G^{\uparrow\downarrow}_{\text{unscaled}} \frac{2\lambda_{\text{sf,Pt}} \tanh(t_{\text{Pt}}/\lambda_{\text{sf,Pt}})}{\sigma_{\text{eff}}}$. Using Pt thickness, \(t_{\text{Pt}} = \SI{0.8}{nm}\)
, spin-diffusion length in Pt, \(\lambda_{\text{sf,Pt}}\approx\SI{0.8}{nm}\) (from our DFT-based hyperbolic fit as explained above), spin Hall angle of Pt, \(|\theta_{\text{SH}}| = 0.007\) (obtained from our DFT calculations for the LSMO/Pt system), Interfacial spin-mixing conductance parameter, \(g^{\uparrow\downarrow} = 1.488 \times 10^{19}\) \si{m^{-2}} (adapted from the experimental work of Gupta \textit{et al.}~\cite{Gupta2021} on a similar LSMO/Pt bilayer) we obtained the value of $\tilde{G}^{\uparrow\downarrow}$. This value is converted to the unscaled conductance \(G^{\uparrow\downarrow}_{\text{unscaled,Re}} = g^{\uparrow\downarrow} (e^2/h) \approx 5.765 \times 10^{14} \, \si{\ohm^{-1}.m^{-2}}\). The imaginary part, \(\text{Im}[G^{\uparrow\downarrow}_{\text{unscaled}}]\), is assumed to be zero for this estimation. To calculate the effective conductivity in the equation for $\tilde{G}^{\uparrow\downarrow}$ for the  \(\sigma_{\text{eff}}\), we utilize the total longitudinal conductivity of our LSMO(2ML)/Pt(4ML) bilayer obtained from our Boltzmann transport calculations, which is \(\sigma_{\text{bilayer}} = \SI{295584.22}{S/m}\). With these factors, we obtain the magnitude of the antidamping torque coefficient to be estimated as $|\tau_{ad}| \approx 0.002 $. 

Therefore, for the ultrathin LSMO/Pt heterostructure, the efficient utilization of the Pt layer thickness relative to its \(\lambda_{\text{sf,Pt}}\) and effective interface spin transmission contribute significantly to the SOT efficiency, compensating to some extent for the reduced effective \(\theta_{\text{SH}}\) of the bilayer.

\begin{figure}
    \centering
\includegraphics[width=0.50\textwidth,height=0.5\textheight,keepaspectratio]{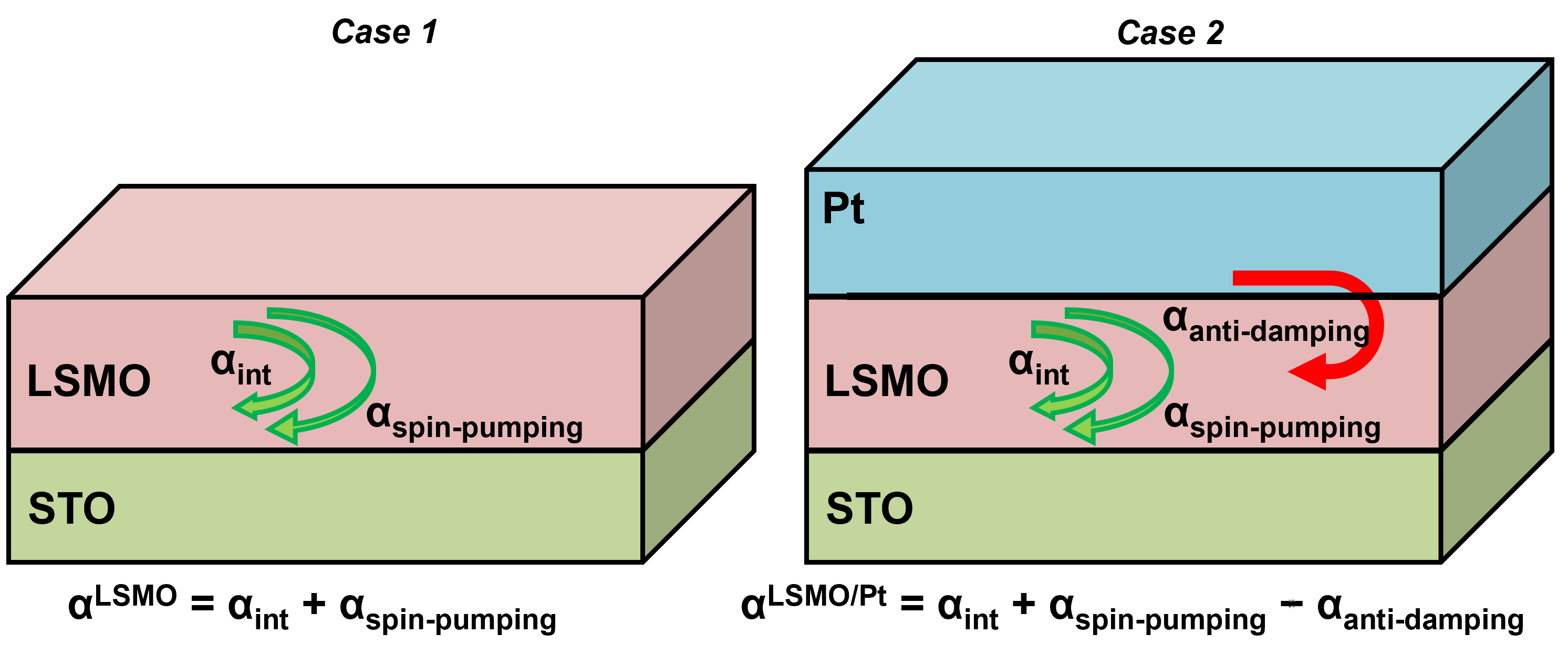}
    \caption{Schematic of damping contributions in LSMO versus LSMO/Pt heterostructures. $\alpha_{int}$ is the intrinsic Gilbert damping that originates from the electronic structure and spin-orbit interaction. $\alpha_{\text{spin-pumping}}$ is contribution due to the ISHE while $\alpha_{\text{anti-damping}}$ originates from the bulk SHE in Pt.}
    \label{TOC}
\end{figure}
\subsection{Contrast with prototypical Co/Pt bilayer}
\begin{figure}
\includegraphics[width=1.0\columnwidth]{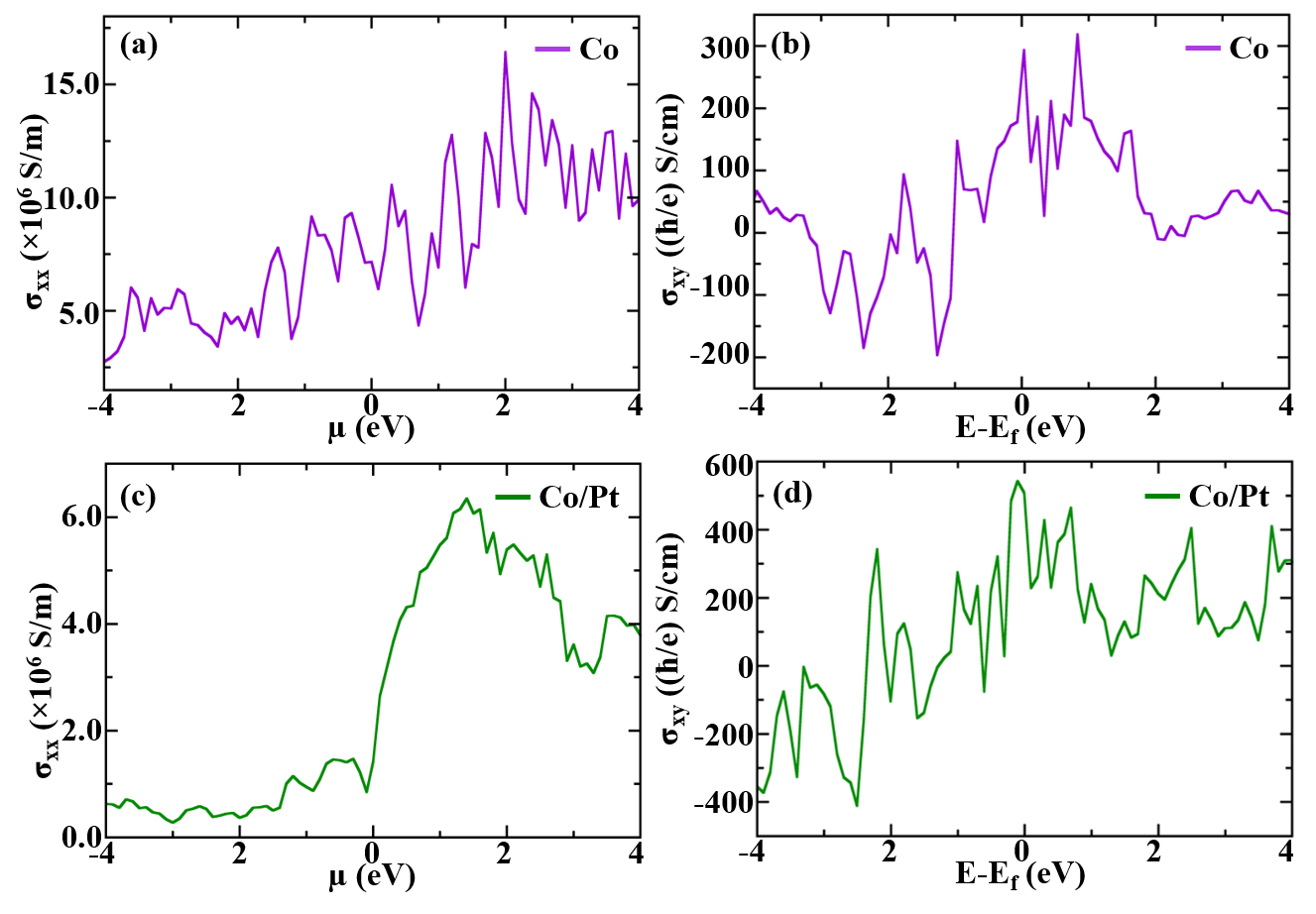}
\caption{The longitudinal charge conductivity and spin Hall conductivities. Upper panels (a\&b):Co, lower panels (c\&d): Co/Pt }
\label{Fig-CoPt}
\end{figure}

Finally, to demonstrate that the above-mentioned results are an exception and not generally expected, we performed similar calculations for the Co/Pt system. The interface was considered using one monolayer of Co and one monolayer of Pt. The observed $\theta_{SH}=0.036$ for the Co/Pt bilayer system is larger than the case of a single Co-monolayer, for which a quite relatively smaller value of $\theta_{SH}=0.004$ is observed. The stronger $\theta_{SH}$ in this heterostructure suggests that the Gilbert damping would increase with increasing Pt thickness due to enhanced spin pumping effect, as for example reported by Mahfouzi \textit{et al.}~\cite{mahfouzi2017intrinsic}. As seen from Fig.\ref{Fig-CoPt}, the Co/Pt interface has smaller longitudinal conductivity and higher SHC in comparison to the Co mono-layer something opposite in nature if one compares with LSMO and LSMO/Pt heterostructure. 
\section{Conclusions}
In summary, the interplay between intrinsic damping mechanisms, spin pumping, and anti-damping torques in FM/HM bilayers presents a rich landscape for theoretical and experimental exploration. Our findings suggest that tailoring the interface properties and electronic structure can significantly impact the damping behavior, offering pathways to optimize materials for spintronic applications. Particularly, this study highlights that it is feasible to engineer materials with desired spin Hall angles (\(\theta_{SH}\)) even in cases where the intrinsic spin-orbit coupling is not prominently strong. This can be achieved by fine-tuning the ratio of spin Hall conductivity (SHC) to longitudinal charge conductivity. By optimizing this ratio, one can effectively control the spin Hall angle, thereby modulating the efficiency of spin-to-charge conversion and the associated damping properties. Future studies could investigate the impact of varying the thickness of the Pt layer, introducing different heavy metals with varying spin-orbit coupling strengths, or applying external fields to modulate the spin dynamics. Understanding the role of temperature and external stress on damping parameters could also provide valuable insights for practical applications.
\section*{Acknowledgments}
This work was supported by the Korea Institute of Science and Technology, GKP (Global Knowledge Platform, Grant number 2V6760) project of the Ministry of Science, ICT and Future Planning.  S.B and P.G.thank the Department of Atomic Energy (DAE) , Govt. of India.

\appendix
\section{Spin-pumping and spin-orbit torque contributions to the effective Gilbert damping}

The effective Gilbert damping in the presence of both spin-pumping and spin-orbit torque (SOT)-induced anti-damping is given by
\begin{equation}
\alpha_{\mathrm{eff}} = \alpha_{\mathrm{int}} + \Delta\alpha_{\mathrm{sp}} - \Delta\alpha_{\mathrm{ad}}.
\label{eq:alpha_eff}
\end{equation}

For the bare LSMO film, the pumped spins are reabsorbed inside LSMO within one spin-diffusion length, whereas for the bilayer they leave the ferromagnet and are absorbed in Pt. Consequently, the spin-pumping mechanisms in the two cases differ.

For the LSMO/Pt interface, spin pumping follows the conventional description~\cite{tserkovnyak2002enhanced}
\begin{equation}
\Delta\alpha_{\mathrm{sp}}^{\text{LSMO/Pt}}
   =\frac{\gamma\hbar}{4\pi M_s\,
       t_{\text{LSMO}}}\,
     g^{\uparrow\downarrow}_{\!\mathrm{eff}},
\label{eq:sp_LSMO_Pt}
\end{equation}
where the heavy metal (Pt) acts as a spin sink. The spin-pumping-induced damping $\Delta\alpha_{\mathrm{sp}}$ depends weakly on the spin Hall angle $\theta_{\mathrm{SH}}$ through spin-backflow correction factor, $\mathcal{R}_{\mathrm{bf}}$,
\begin{equation}
\Delta\alpha_{\mathrm{sp}} = \frac{\gamma\hbar}{4\pi M_s t_{\mathrm{FM}}} g^{\uparrow\downarrow}_{\mathrm{eff}}, 
\quad g^{\uparrow\downarrow}_{\mathrm{eff}}=g^{\uparrow\downarrow}(1-\mathcal{R}_{\mathrm{bf}}), 
\quad \mathcal{R}_{\mathrm{bf}}\propto\theta_{\mathrm{SH}}^2.
\label{eq:sp_backflow}
\end{equation}

For a bare LSMO film, there is no spin sink; the angular momentum must be absorbed within the material. Following Gupta \textit{et al.}~\cite{gupta2024self}, $\alpha_{\mathrm{eff}}=\alpha_{\mathrm{int}}+\alpha_{\mathrm{sp}}\propto\sqrt{\theta_{\mathrm{SH}}}$, which can be approximated as
\begin{equation}
\alpha \approx \sqrt{\frac{\theta_{\mathrm{SH}}\, w(2 e / \hbar) \lambda_{LSMO} \tanh \!\left( \frac{t}{2 \lambda_{LSMO}} \right)}{I_{\mathrm{ISHE}}} \frac{g_r^{\uparrow,\downarrow} \gamma B_{\mathrm{rf}}^2 \hbar}{4 \pi \mu_0 M_s}}.
\label{eq:alpha_self_pumping}
\end{equation}
Here $I_{\mathrm{ISHE}}$ is the measured inverse spin Hall current and $\lambda_{LSMO}$ is the spin-diffusion length in the LSMO film. 

While the above approach is more physically motivated, it requires accurate transferability between computationally obtained parameters and experimental conditions. Moreover, certain parameters like $B_{rf}$ (the rf magnetic field associated with microwaves) are experimental in nature and challenging to assume from the theory.

A simpler alternative is to use Eq.~(\ref{eq:sp_LSMO_Pt}) with a different length scale. In the self-pumping case, the angular momentum is absorbed within the ferromagnet, and we may introduce an \textit{effective absorption depth} $\lambda_{\mathrm{abs}}$ equal to the measured spin-diffusion length in LSMO ($\approx0.8$~nm) determined from hot-electron attenuation spectroscopy~\cite{rana2013hot}:
\begin{equation}
\Delta\alpha_{\mathrm{sp}}^{\text{LSMO}}
   =\frac{\gamma\hbar}{4\pi M_s\,
       \lambda_{\text{abs}}}\,
     g^{\uparrow\downarrow}_{\!\mathrm{eff}}.
\label{eq:sp_self}
\end{equation}

To calculate the anti-damping contribution numerically, we may start with the interfacial spin–orbit torque density localized at the FM$|$HM interface ($z=0$)~\cite{haney2013current},
\begin{equation}
\mathbf{T}(z)=\delta(z)\,\frac{g \mu_B\, J_c}{2 e}\left[\tau_{ad}\, \hat{\mathbf{m}} \times(\hat{\mathbf{m}} \times \hat{\mathbf{y}})+\tau_f\, \hat{\mathbf{m}} \times \hat{\mathbf{y}}\right],
\end{equation}
where $J_c$ is the in-layer charge current density in the heavy metal, and $\tau_{{ad},f}$ are the dimensionless (anti)damping-like and field-like torque efficiencies, respectively. For a ferromagnet of thickness $t_F$, the spatially averaged torque density is $\overline{\mathbf{T}}=\mathbf{T}_{\rm int}/t_F$. Balancing this against the LLG torque $-\gamma \mu_0 M_s\,\hat{\mathbf m}\times(\hat{\mathbf m}\times \mathbf{H}_{AD})$ and using $g\mu_B=\hbar\gamma$  gives the effective \emph{anti-damping (damping-like) field}
\begin{equation}
\boxed{\;\mu_0 H_{AD} \;=\; \frac{\hbar}{2e}\,\frac{J_c\,\tau_d}{M_s\,t_F}\; } .
\end{equation}
The damping-like efficiency we have already mentioned in the main text ($\tau_{ad}=0.002$). 

To express this as a scalar change in the Gilbert damping, we map the anti-damping field to a linewidth-equivalent increment under small-angle steady precession at angular frequency $\omega$,
\begin{equation}
\boxed{\Delta \alpha_{ad}\;\approx\; \frac{\gamma\,\mu_0 H_{AD}}{\omega}\,\cos\varphi\;} ,
\end{equation}
where $\varphi$ is the phase between the SOT drive and $\dot{\hat{\mathbf m}}$ set by the microwave circuit and device geometry. In the absence of an independent phase calibration, taking $\varphi=0$ (i.e.\ $\cos\varphi=1$) provides an \emph{upper-bound} estimate. For typical  measurements one may use $f=\omega/2\pi = 10~\text{GHz}$. 

Using realistic experimental parameters, the calculated contributions to the total damping are summarized in Table~\ref{tab:alpha}. While the theoretical values do not exactly match the experiment, they reproduce the qualitative trends.
\begin{table}[h]
\centering
\caption{Calculated contributions to the damping and comparison with experiment (f=10~GHz, $J_c=10^{11}$~A/m$^2$).}
\begin{ruledtabular}
\begin{tabular}{lccccc}
Sample & $\theta_{\mathrm{SH}}$ & $\alpha_{\mathrm{int}}$ & $\Delta\alpha_{\mathrm{sp}}$ & $-\Delta\alpha_{\mathrm{ad}}$ & $\alpha_{\mathrm{eff}}$ (calc./exp.) \\
\hline
LSMO  & 0.093 & 0.0014 & 0.0788 & 0 & 0.0802 / 0.0104 \\
LSMO/Pt & 0.007 & 0.012 & 0.0064 & 0.005 & 0.0134 / 0.0046 \\
\end{tabular}
\end{ruledtabular}
\label{tab:alpha}
\end{table}

In the calculations, we used the experimental $g^{\uparrow\downarrow}$, with $\theta_{\mathrm{SH}}$, $M_s$, and $\lambda_{\mathrm{sf}}$ obtained from our \textit{ab initio} calculations.

\clearpage
\bibliography{SHC}
\bibliographystyle{apsrev4-2}
\end{document}